# Quantitative measurement of cooperative binding in partially-dissociated water dimers at the hematite "R-cut" surface

Paul T. P. Ryan[1], Panukorn Sombut[1], Ali Rafsanjani Abbasi[1], Chunlei Wang[1], Fulden Eratam[2], Francesco Goto[2,3], Ulrike Diebold[1], Matthias Meier[1,4], David A. Duncan[2], Gareth S. Parkinson[1]

[1]Institute of Applied Physics, Technische Universität Wien, Vienna, Austria
[2]Diamond Light Source, Harwell Science and Innovation Campus, Didcot, UK
[3]Politecnico di Milano, Piazza Leonardo da Vinci, Milano MI, Italy
[4]Faculty of Physics and Center for Computational Materials Science, Uni. of Vienna, Vienna, Austria

**Abstract**
Water-solid interfaces pervade the natural environment and modern technology. On some surfaces, water-water interactions induce the formation of partially-dissociated interfacial layers; understanding why is important to model processes in catalysis or mineralogy. The complexity of the partially-dissociated structures often make it difficult to probe them in a quantitative manner. Here, we utilize normal incidence x-ray standing waves (NIXSW) to study the structure of partially-dissociated water dimers ($H_2O$-OH) at the α-$Fe_2O_3$(012) surface (also called ($1\bar{1}02$) or "R-cut" surface); a system simple enough to be tractable, yet complex enough to capture the essential physics. We find the $H_2O$ and terminal OH groups to be the same height above the surface within experimental error (1.45 $\pm$ 0.04 Å and 1.47 $\pm$ 0.02 Å, respectively), in line with DFT-based calculations that predict comparable Fe-O bond lengths for both water and OH species. This result is understood in the context of cooperative binding, where the formation of the H-bond between adsorbed $H_2O$ and OH induces the $H_2O$ to bind more strongly, and OH to bind more weakly compared to when these species are isolated on the surface. The surface OH formed by the liberated proton is found to be in plane with a bulk truncated (012) surface ($-0.01 \pm 0.02$ Å). DFT calculations based on various functionals correctly model the cooperative effect, but overestimate the water-surface interaction.

**Introduction**

Metal oxide surfaces are omnipresent in the environment, and their interaction with water underlies natural processes such as geochemistry, corrosion and cloud formation. Metal-oxides are also often employed as catalysts, catalytic supports, and electrocatalysts, and it is known that adsorbed water affects the catalytic process even in cases where it is not directly a reactant.[1] For example, water adsorbed at the oxide surface is known to affect the morphology and reactivity of supported metal adatoms or clusters,[2–5] and metal-oxides utilized as electrocatalysts can undergo hydroxylation and oxygen exchange reactions with the water.[6–9] Correctly modelling the water-oxide interaction is therefore an important issue, and a prerequisite for understanding how these materials behave under realistic application conditions.

Hematite (α-$Fe_2O_3$) is a naturally abundant mineral which has shown promising potential in the context of photochemical water splitting. It has a 2 eV bandgap, which facilitates oxygen evolution using visible light.[10–12] Recently, hematite has found use as a support for so-called single-atom catalysts for reactions including the water-gas shift reaction and the (electrochemical) oxygen reduction reaction.[13–16] The α-$Fe_2O_3$(012)-(1×1) surface (also called ($1\bar{1}02$) or "R-cut" surface) is one of the most prevalent low index facets, and water adsorption has been studied previously in both UHV[17–19] and in liquid[6,20,21]. All studies to date suggest that water exposure leads to both molecular and dissociated components at the interface. Recently, we studied water adsorption on this surface[20] using non-contact atomic force microscopy (nc-AFM), x-ray photoelectron spectroscopy (XPS) and density functional theory (DFT)-based calculations, and concluded that the surface stabilizes $H_2O$-OH dimers. More specifically, it was found that isolated $H_2O$ molecules adsorb intact at a surface cation, but the interaction with a second $H_2O$ leads to its dissociation into a terminal hydroxyl ($OH_t$) adsorbing at a neighboring cation site and an additional surface hydroxyl ($OH_s$) species at a lattice oxygen on the surface. The $H_2O$ and $OH_t$ form a hydrogen bond leading to a partially-dissociated dimer ($H_2O$-$OH_t$) as shown schematically in Figure 1 b and c.

The phenomenon of partial dissociation has been reported previously for several metal oxide[22–28] and metal[29–31] surfaces. It occurs when the energy gained through the formation of a $H_2O$-$OH$ hydrogen bond compensates the energy lost creating the less favorable adsorbate (in isolation). The interaction can be further strengthened by a so-called "cooperative binding" effect [32–34], in which it is assumed that water molecules optimally donate and receive equally in their bonding interactions. Thus, a stronger intermolecular hydrogen bond is accompanied by a stronger surface bond between the water and the surface, which should manifest as shorter water/cation bond lengths. This is the effect that we aimed to directly measure in the current manuscript.

The structure of the substrate is important for observing partial dissociation and cooperative binding, because undercoordinated cation-anion pairs are required to host the $H_2O$ and OH groups. Also, the cation-cation distance must be short enough to facilitate the formation of a strong hydrogen bond between molecular $H_2O$ and the terminal $OH_t$. Complicated arrangements can occur on surfaces where the $H_2O$ and OH groups form overlayers with large unit cells,[22,23] which makes elucidation of the structure challenging. DFT-based calculations are often utilized, but modelling such systems is difficult due to the subtle balance of the interactions involved and the need to account for dispersion interactions. In contrast, the $H_2O$/α-$Fe_2O_3$(012)-(1×1) system is unusual in that it limits the size of partially dissociated agglomerates to $H_2O$-$HO_t$ pairs, offering a comparatively simple system for interrogation of the cooperative binding effect.

In this study we utilize the quantitative structural technique normal incidence x-ray standing waves (NIXSW), to chemically resolve the adsorption sites of species in the $H_2O$-OH dimer on α-$Fe_2O_3$(012)-(1×1). Our results show that the $H_2O$ and $OH_t$ groups reside at the same height above the surface (1.45 ± 0.04 Å and 1.47 ± 0.02 Å, respectively), which implies a similar Fe-O bond length for the $H_2O$ and $OH_t$ (≈ 2.0 Å). These results agree with the results of our prior DFT calculations, and corroborate the picture of partially-dissociated water dimers originally derived from nc-AFM images. Also, they provide compelling evidence for the involvement of cooperative binding effects for water adsorption on metal oxide surfaces. Such observations are found in prior studies,[22–27] though to our knowledge our results represent the first quantitative evidence of this cooperative binding effect.

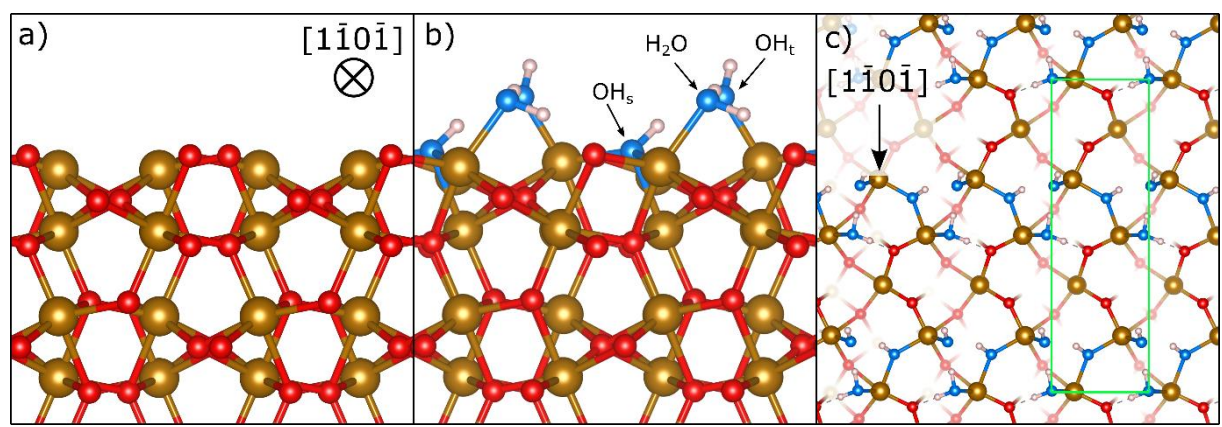

Figure 1 – Model for water adsorption on the α-$Fe_2O_3$(012)-(1×1) surface.[20] a) Side view of the clean bulk truncated α-$Fe_2O_3$(012)-(1×1) surface. b) side view of the DFT model showing the $H_2O$-$OH_t$ dimer and $OH_s$. The surface atoms have negligible vertical relaxation. c) top view of the DFT model with a green (1×3) unit cell. Red and brown atoms are O anions and Fe cations of the bulk, respectively. Blue atoms are the oxygens of the $H_2O$, $OH_t$ and $OH_s$ species and white atoms are hydrogens.

## Experiment and Methods

### Samples

A polished α-Fe₂O₃(012) "R-cut" surface single crystal (± 0.1°, from the SurfaceNet GmbH) was prepared in situ via several cycles of sputtering (Ar⁺, voltage: 1 keV, emission current: 3 μA, 30 min) and annealing in $2 \times 10^{-6}$ mbar of oxygen (~500 °C, 30 min). The prepared samples showed a (1×1) LEED pattern consistent with a bulk-truncated surface.[18,19] A side view of this surface structure is shown in Figure 1 a.

High purity deionized water was obtained from a Milli-Q system and cleaned in-situ by several freeze-pump-thaw cycles. The clean α-Fe₂O₃(012)-(1×1) surface was exposed to $3 \times 10^{-8}$ mbar of water for 10 minutes at 300 K (~14 L, where 1 L is $1 \times 10^{-6}$ mbar·s). Following transfer into the analysis chamber (within 5-10 minutes), the sample was cooled to 200 K using a liquid nitrogen cryostat. Based on our previous study, this preparation procedure should produce a (1×3) overlayer of partially-dissociated water dimers, as shown in Figure 1 b and c.[20]

### NIXSW and SXPS

The NIXSW technique exploits the x-ray standing wave formed by the interference between the incident and reflected waves around the Bragg condition for a given reflection $(h, k, l)$.[35–37] The standing wave's period matches the interplanar spacing $d_{hkl}$ between the Bragg diffraction planes.[38] The standing wave's phase, and thus the location of its maximum intensity, varies as the photon energy is scanned through the Bragg condition. When the phase is π, the maximum intensity is halfway between Bragg diffraction planes; when the phase is zero the maximum intensity is coincident with the Bragg diffraction planes. Any atom within this standing wavefield will therefore experience a varying electromagnetic field intensity as a function of its position between these Bragg diffraction planes. This variation in intensity results in a characteristic absorption profile, which can be acquired by monitoring the relative photoelectron yield. The measured profile is then fitted uniquely, using dynamical diffraction theory,[39] by two dimensionless parameters: the coherent fraction, $f_{hkl}$, and the coherent position, $p_{hkl}$. These can be considered to correspond to the degree of order and the mean position of the absorber atoms relative to the Bragg diffraction planes, respectively.[36,37] When the origin of the substrate atomic coordinates is chosen to be in the surface plane, the coherent position is related to the mean adsorption height ($H$) by:

$$H = (n + p_{hkl} - 0.61) \cdot d_{hkl}, \tag{1}$$

where $d_{hkl}$ is the reflection layer spacing and $n$ is an integer which relates to so called "modulo-d" ambiguity,[36] where adsorption heights that differ by the interplanar spacing cannot be directly differentiated. In practice, however, the correct value of $n$ can often be easily assigned as $d_{hkl}$ typically is in the order of ~2 Å, thus it is generally trivial to exclude adsorption heights that are unphysically low or high. Since we only utilize the (024) reflection here, $d_{024} = 1.84$ Å and the coherent fraction and coherent position are denoted as: $f_{024}$ and $p_{024}$. Note that, due to the standing wave being generated by the crystallinity of the bulk substrate, the adsorption height measured in NIXSW is not relative to the position of the outermost atoms at the surface, but rather to a projected bulk-like termination of the surface. To obtain adsorption heights relative to the bulk-like surface O atoms, the coherent position of the surface O atoms (0.61), has been subtracted from $p_{hkl}$ in equation (1). Our DFT calculations indicate that the terminal O atoms have a negligible vertical surface relaxation after the adsorption of water (see Table 1), thus making $H$ a direct measure of the adsorption height with respect to the surface oxidic O atoms.

By acquiring the photoelectron yield from the O 1s core level as a probe of the NIXSW absorption rate, we obtain a chemically resolved probe that permits signals from the bulk oxide, surface hydroxide and adsorbed water to be discriminated independently.

All measurements were conducted in the permanent ultra-high vacuum (UHV – ~$1 \times 10^{-10}$ mbar) end station on the I09 beamline[40] at the Diamond Light Source. Beamline I09 utilises two separate undulators, which are monochromated separately by a double Si(111) crystal monochromator and a plane grating monochromator. These two separate lines provide simultaneous access to both 'hard' and 'soft' x-ray energies, respectively. Specifically, we have used incident photon energies of 650 eV for all the O 1s soft x-ray photoelectron spectroscopy (SXPS) measurements. For the NIXSW (024) reflection, a hard photon energy range of 3350–3370 eV was used. The absolute binding energy scale of all XP spectra were calibrated by subsequent measurements of the Au 4f core-level from a gold foil situated below the sample holder.

All photoelectron spectra were acquired using a VG Scienta EW4000 HAXPES hemispherical electron analyser (angular acceptance range ± 28°) mounted perpendicular to the incident radiation and in plane with the polarisation of the incident photon (linear-horizontal). All photoelectron spectra were peak fitted using a numerical convolution of a Lorentzian and Gaussian peak profile. For all peaks in all spectra, the same Lorentzian peak width was used, as determined from the fits of bulk oxide O 1s photoemission peak, and the Gaussian peak width was allowed to vary. A Shirley background[41] was subtracted from each spectrum.

**Theoretical details**

We utilized the Vienna *ab initio* Simulation Package (VASP)[42,43] with the optB88-DF[44,45] functional utilizing a $U_{\text{eff}}$ of 5 eV[46]. Additionally, we investigated hybrid functional (HSE06) with the fraction of exact exchange of 12% and 25% and a range separation parameter of $0.2^{-1}$ Å$^{-1}$. A further set functionals were tested (in total 20) and details and citations for these other functionals are found in Section 1 of the ESI.

The surface calculations employed symmetric slabs, with only the two inner central O layers kept fixed. The model of a (1×3) overlayer of partially-dissociated water ($H_2O$-OH) dimers contains four water molecules per (1×3) unit cell (Figure 1 b and c).[20] Two $H_2O$ molecules are molecularly adsorbed on Fe cation sites, two $H_2O$ molecules dissociate, liberating two protons to two surface oxygen atoms to form two OH$_s$ species in the O surface plane and two OH$_t$ species terminal to surface Fe cations. This model is derived from out prior nc-AFM experimental study.[20] Between neighboring $H_2O$-OH$_t$ dimers along the $(0\bar{1}0\bar{1})$ direction, one Fe cation site is left vacant. Note that the adsorption site atop a surface Fe cation is where the next O atoms would reside if the bulk corundum structure were continued outward, although in that case the next layer of the bulk structure would have a larger height (1.62 Å) than that found for OH$_t$ / $H_2O$ (~1.46 Å). Further details of the computational setup are provided in the Section 1 of the ESI.

The average adsorption energy per $H_2O$ molecule ($E_{ads}$) is computed according to the formula:

$$E_{ads} = \left(E_{Fe_2O_3+nH_2O} - \left(E_{Fe_2O_3} + nE_{H_2O}\right)\right)/n, \tag{2}$$

where $E_{Fe_2O_3+nH_2O}$ is the total energy of the α-Fe$_2$O$_3$(012) surface with adsorbed $H_2O$, $E_{Fe_2O_3}$ is the total energy of the clean α-Fe$_2$O$_3$(012) surface, the $E_{H_2O}$ represents the energy of the $H_2O$ molecule in the gas phase, and $n$ is the number of $H_2O$ molecules.

The O 1s core-level binding energies are calculated in the final state approximation.[47] The calculation was undertaken with respect to the oxygen in the bulk position.

To elucidate the underlying cooperative binding mechanism, we calculated the total variation of charge transfer ($\Delta_{Tot}$) between the on-the-surface adsorbed $H_2O$-$OH_t$ dimer and the on-the-surface adsorbed individual molecules. $\Delta_{Tot}$ is defined as:

$$\Delta_{Tot} = \Delta_{dimer} - (\Delta_{H2O} + \Delta_{OHt+OHs}), \qquad (3)$$

with:

$$\Delta_{dimer} = \rho_{dimer} - \rho_{surface}, \qquad (4)$$

$$\Delta_{H2O} = \rho_{H2O} - \rho_{surface}, \qquad (5)$$

$$\Delta_{OHt+OHs} = \rho_{OHt+OHs} - \rho_{surface}, \qquad (6)$$

where $\rho_x$ is the electronic charge distribution given by DFT for configuration $x$. $\Delta_{Tot}$ can thus be recast as:

$$\Delta_{Tot} = \rho_{dimer} - \rho_{H2O} - \rho_{OHt+OHs} + \rho_{surface}, \qquad (7)$$

where the surface is needed once to compensate for the double counting of it in the individual-molecule interactions. A positive value can therefore be attributed to a stronger bonding between the molecules involved in the dimer. The individual-molecule components are calculated at partially dissociated water dimer positions (no relaxations allowed) but with the counter component in the dimer removed.

## Results

### SXPS

Figure 2a shows the O 1s SXPS spectra for the clean α-$Fe_2O_3$(012)-(1×1) surface and for the surface after exposure to 14 L of water vapour at 300 K (measured at 200 K). Upon exposure to water, two new photoemission peaks are visible at higher binding energies than the main bulk oxide peak. The bulk O 1s peak is also found to shift to higher binding energies. This has been observed previously and is attributed to band bending.[20] Figure 2b shows the peak fitted spectrum from Figure 2a after exposure to water. Three peaks were used in the fitting and assigned as oxygen from the bulk oxide, $O_{bulk}$, oxygen from adsorbed hydroxides, $O_{OH}$, and oxygen from adsorbed water, $O_{H2O}$.

The binding energies of the $O_{H2O}$ and $O_{OH}$ peaks correspond well to prior XPS studies of adsorbed water and hydroxides on metal-oxides.[22,48–51] Our DFT calculations also show that the $OH_t$ and $OH_s$ species have a O 1s core level binding energy that differs by only ~ 0.1 eV, which is within the error of the calculation. Comparison of the $O_{H2O}$ and $O_{OH}$ relative peak areas shows that there is 21% more OH than expected. From the model of our prior nc-AFM study[20] one would expect exactly double $O_{OH}$ vs $O_{H2O}$. The 21% increase in OH population is most likely due to extra dissociative adsorption at defects or step edges, as no beam damage was observed during the measurements.

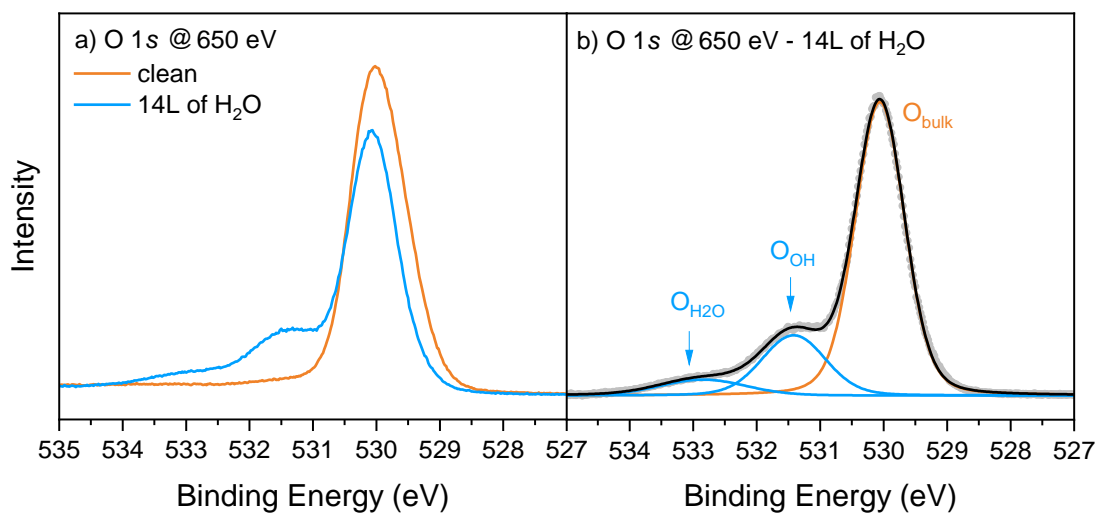

Figure 2 – The SXPS results. a) O 1s SXPS core-level spectrum of the clean α-Fe$_2$O$_3$(012)-(1×1) surface and after exposing this surface to 14L of H$_2$O. b) The same spectrum from a) after dosing H$_2$O but peak fitted with three components assigned as oxygen from the bulk crystal, O$_{bulk}$, adsorbed hydroxyls, O$_{OH}$, and adsorbed water, O$_{H2O}$.

**NIXSW**

Figure 3 shows the NIXSW photoelectron yield profiles for the O$_{OH}$ and O$_{H2O}$ oxygen, as well as the measured intensity of the (024) reflection. The fitted coherent fraction of the O$_{H2O}$ photoemission peak ($f_{024} = 0.91 \pm 0.04$) implies that the H$_2$O occupies an extremely well-defined position normal to the (012) surface with this position defined by the fitted coherent position ($p_{024} = 0.40 \pm 0.02$). The small deviation from unity in the coherent fraction can be attributed solely to molecular and crystal vibrations.[52]

As it is assumed that the H$_2$O adsorbs above the surface, only values of $n > 1$ in equation 1 are considered for the H$_2$O. Should $n \geq 2$ be considered the adsorption height of the H$_2$O would be unphysically large (> 3 Å). Thus, it is assumed that $n = 1$ and the adsorption height of H$_2$O above the bulk-like terminated α-Fe$_2$O$_3$(012) surface is $H_{H2O} = 1.45 \pm 0.04$ Å close to a bulk continuation adsorption height of 1.61 Å. This is shown schematically in Figure 4.

The fitted coherent fraction of the O$_{OH}$ photoemission peak ($f_{024} = 0.59 \pm 0.02$) is significantly lower than that of O$_{H2O}$. Such a low coherent fraction (< 0.8) cannot be attributed to molecular or crystal vibrations alone,[52] so the O$_{OH}$ photoemission peak must correspond to chemically similar oxygen atoms located at different distinct heights in the [024] direction. This makes sense, because water dissociation is expected to produce both a OH$_t$ adsorbed above a surface cation, and a OH$_s$ at a surface oxygen atom.[20]

It is possible to extract the two individual OH adsorption heights by making reasonable assumptions about OH$_s$ and OH$_t$. We assume that the 21 % excess of O$_{OH}$ signal contributes decoherently to the OH position giving an order parameter [53] of $C = 0.79$ (see ESI Section 2 for further explanation). This would be the case if some molecules adsorb dissociatively at defects and step edges. Also, we assume a similar reduction of $f_{024}$ from thermal vibrations as found for the O$_{H2O}$ species (Debye-Waller factor = 0.91). Both of these factors lead to a true structural $f_{024} = 0.82$.[53] Finally, by assuming an equal occupation of OH$_t$ and OH$_s$, the analysis of the fitted coherent position ($p_{024} = 0.51 \pm 0.01$) leads to a OH$_t$ coherent position of $p_{024} = 0.41 \pm 0.01$ and a OH$_s$ coherent position of $p_{024} = 0.62 \pm 0.01$.

A more detailed description of the calculation is provided in Section 2 of the ESI. The results from other models, varying the relative population of the OH$_t$ and OH$_s$ sites, the order parameter C and the Deby-Waller factor are also provided in Section 2 of the ESI (Models 1-4). Generally, the physically reasonable models in the ESI have results close to the model presented here within the main article (Model 2). While there may be some ambiguity on the precise adsorption height of both $OH_s$ and $OH_t$, all of the models that are not excluded by our prior nc-AFM measurements, indicate that the $OH_s$ species is effectively in plane with the surface oxygen atoms and that the $OH_t$ species is effectively in plane with the water molecule at the approximate height of the O bulk continuation site.

Using equation (1), we can calculate heights for the two OH species with respect to a (012) bulk truncated oxygen surface layer for Model 2. Following the same arguments as for the $H_2O$, n = 1 for the $OH_t$ and n = 0 for the $OH_s$. In turn, we find that the $OH_t$ sits at a height $H_{OHt} = 1.47 \pm 0.02$ Å above the surface; in plane with the adsorbed $H_2O$. The $OH_s$ is found in plane with the surface oxygens ($H_{OHs} = -0.01 \pm 0.02$ Å). These heights are shown schematically in Figure 4. Note that the lateral placement of the species is not determined here and for the schematic has been defined as oxygen bulk (continuation) sites.

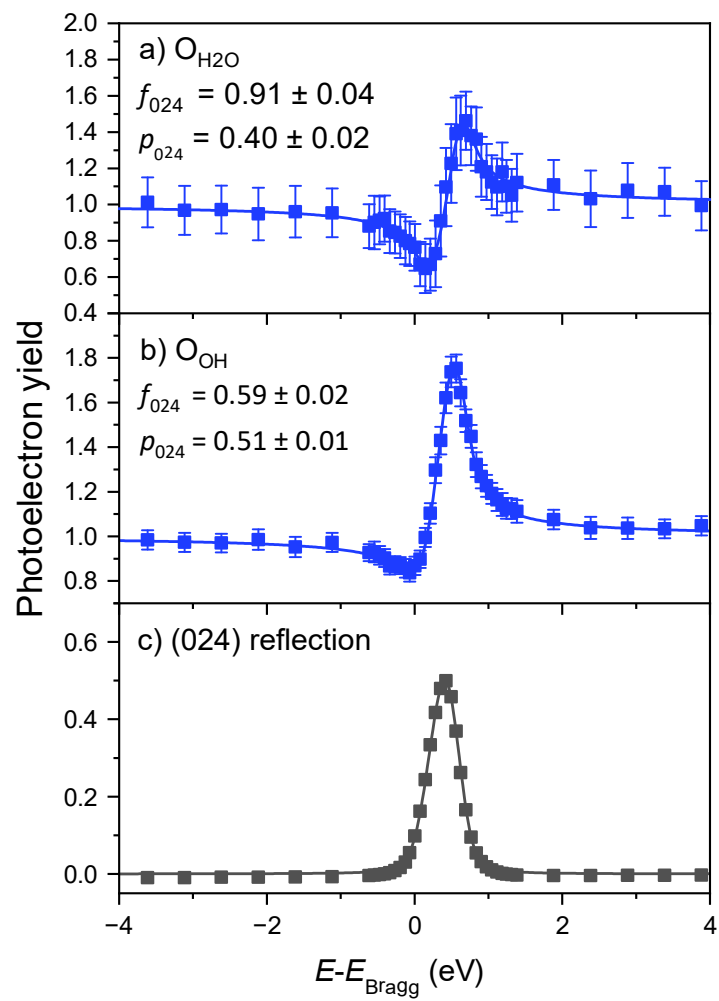

Figure 3 – The NIXSW results. Photoelectron yield profiles for the a) $O_{H2O}$ and b) $O_{OH}$ photoemission peaks and c) the intensity of the (024) reflection. Given are the fitted coherent fraction, $f_{024}$, and coherent position, $p_{024}$, for each absorption profile.

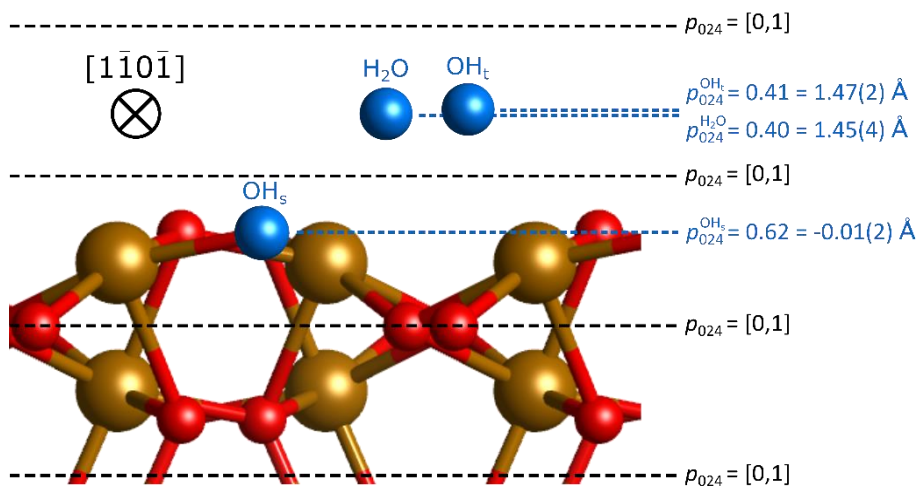

Figure 4 – Schematic showing the positions of the H₂O, OH_t, and OH_s in blue projected onto a bulk truncated (012) surface. The (024) reflection periodicity is shown by the black dashed lines. Given in blue are the measured or calculated coherent positions, $p_{024}$, for each species and the corresponding heights above a (012) bulk oxygen surface layer. In brackets is the error. Note that this schematic does not take into account the lateral placement of the H₂O and OH_t and this has been chosen as oxygen bulk (continuation) sites.

## DFT

Table 1 compares the results of the DFT calculations for the H₂O-OH_t dimer with the measured experimental adsorption heights ($H_{H2O}$, $H_{OHt}$ and $H_{OHs}$) from the NIXSW. These heights were calculated in the same manner in which the NIXSW measurements are undertaken, by projecting the species' position onto a bulk unit cell. This ensures that any bulk or surface relaxations are taken into account and, because the NIXSW measurement is not sensitive to any bulk or surface relaxations,[37] this allows for direct comparison with the NIXSW results.

The optB88-DF, HSE 12% and 25% functionals were selected for comparison to the NIXSW results because they reproduce the bulk lattice parameters of the α-Fe₂O₃ crystal extremely well. This is demonstrated by Δc in Table 1 which is the difference between the experimental and optimized c lattice parameter for each functional. Note that in the actual calculations presented here, the experimental lattice parameters were used so that comparison to the NIXSW measurements could be made. We have also calculated the same results for other functionals and these are given in Section 3 of the ESI along with further detailed discussions of all the theoretical results. By undertaking correlational analysis with all the tested functionals, we find that all the structural values (i.e. both heights and bond lengths) correlate extremely strongly with the Δc parameter (correlation coefficients > 0.9, Table S3). This reinforces our choice of optB88-DF, HSE 12% and 25% calculations as being the most suited for comparison with the NIXSW results.

In terms of $H_{H2O}$, all three functionals overbind the position of the H₂O (Table 1). However, we believe the OptB88-DF functionals performs best overall, placing the H₂O and OH_t almost coplanar as determined by the NIXSW measurements. The hybrid functionals perform poorly in this regard, placing the OH_t consistently too high with respect to the H₂O species. In all cases, both the H₂O and OH_t adsorption heights and Fe-O bond lengths are of similar order, between 1.4-1.5 Å (for heights) and 1.9-2.1 Å (for bond lengths; Table 2). Our DFT calculations also show that vertical relaxations of the surface Fe atoms bound to the H₂O and OH_t species, $\Delta H_{FeH2O}$ and $\Delta H_{FeOH}$ (Table 1), are at most ≈ + 0.05 Å and ≈ + 0.1 Å respectively. These relaxations would imply very similar bond lengths for the Fe-H₂O and Fe-OH_t bonds and this is indeed the case (Table 2); the + 0.1 Å upwards relaxation of the Fe bound to OH_t compensates for the higher position of the OH_t.

Figure 5 shows isodensities of the calculated $\Delta_{Tot}$ which represents the change in charge between isolated H₂O and OH_t + OH_s vs the H₂O-OH_t dimer. Yellow isosurfaces depict a reduction in charge in the dimer case, when compared to isolated species. Cyan isosurfaces depict an increase in charge in the dimer case. In general, charge is found to be reorganized away from the Fe-OH$_t$ bond and towards the Fe-H$_2$O bond via a hydrogen bond in the dimer. This explains the observed changes in Fe-O bond lengths.

Table 1 – The DFT heights ($H_{H2O}$, $H_{OHt}$ and $H_{OHs}$) of the species in the H$_2$O-OH$_t$ dimer with comparison to the NIXSW results of Model 2. For the H$_{OHt}$ and H$_{OHs}$ results the range of values for Models 1-4 (see Section 2 of the ESI) are also given. These DFT heights are calculated with respect to an oxygen bulk terminated (012) surface. Values in brackets are the error in the last significant figure. $\Delta H_{Os}$ is average change in vertical height of the surface oxygens after water exposure. $\Delta H_{FeH2O}$ and $\Delta H_{FeOH}$ are the vertical height changes of the Fe atoms bound to the H$_2$O and OH$_t$ species (compared to dry bulk surface). E$_{ads}$ is the calculated adsorption energy of the H$_2$O. $\Delta c$ is the difference between the experimental and calculated c unit cell parameter for α-Fe$_2$O$_3$.

|  | $H_{H2O}$ (Å) | $H_{OHt}$ (Å) | $H_{OHs}$ (Å) | $\Delta H_{Os}$ (Å) | $\Delta H_{FeH2O}$ (Å) | $\Delta H_{FeOH}$ (Å) | $E_{ads}$ (eV) | $\Delta c$ (Å) |
|---|---|---|---|---|---|---|---|---|
| Model 2 | 1.45(4) | 1.47(2) | -0.01(2) | - | - | - | - | - |
| Models (1-4) | - | 1.38-1.47 | -0.01-0.09 | - | - | - | - | - |
| HSE 12% | 1.42 | 1.50 | 0.11 | 0.00 | + 0.06 | + 0.10 | -1.19 | 0.06 |
| HSE 25% | 1.36 | 1.44 | 0.06 | -0.04 | + 0.02 | + 0.10 | -1.30 | -0.01 |
| OptB88-DF | 1.42 | 1.47 | 0.11 | -0.02 | + 0.04 | + 0.10 | -1.60 | 0.02 |

Table 2 – The DFT Fe-O bond lengths for the species in the H$_2$O-OH$_t$ dimer ($d_{Fe-H2O}$ and $d_{Fe-OHt}$) and for isolated species ($d^{iso}_{Fe-H2O}$ and $d^{iso}_{Fe-OHt}$) on the α-Fe$_2$O$_3$(012)-(1×1) surface. The cooperative binding trend is seen in all DFT functionals. The Fe-H$_2$O bond lengths are shorter in the dimer when compared to the isolated species and vice versa for the OH$_t$.

|  | $d_{Fe-H2O}$ (Å) | $d_{Fe-OHt}$ (Å) | $d^{iso}_{Fe-H2O}$ (Å) | $d^{iso}_{Fe-OHt}$ (Å) |
|---|---|---|---|---|
| HSE 12% | 2.08 | 1.94 | 2.12 | 1.88 |
| HSE 25% | 2.07 | 1.93 | 2.11 | 1.88 |
| OptB88-DF | 2.07 | 1.95 | 2.10 | 1.90 |

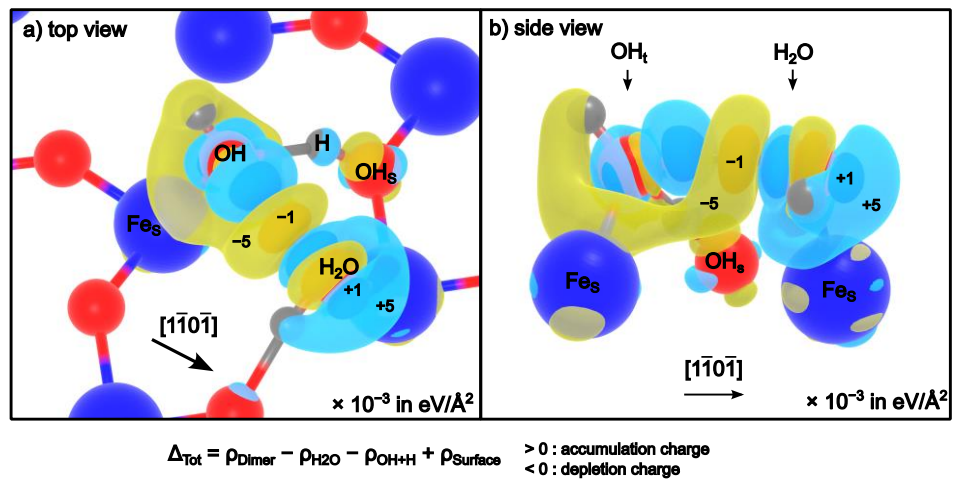

Figure 5 – The total variation of charge transfer ($\Delta_{Tot}$). a) Top view of the α-Fe$_2$O$_3$(012)-(1×1) surface with isosurfaces of the calculated $\Delta_{Tot}$ overlaid. In short, yellow depicts a reduction of charge and cyan depicts an increase in charge when going from isolated OH$_t$ and H$_2$O species to a H$_2$O-OH$_t$ dimer. The numbers quantify this change, being values of $\Delta_{Tot} \times 10^{-3} eV/Å^2$ (equation 6). Charge is found to be reorganized to the Fe-H$_2$O bond and reorganized away from the Fe-OH$_t$ bond which explains the experimentally observed shortening and lengthening of these bonds respectively. b) A side view of the surface. Red spheres are O atoms, blue spheres are Fe atoms.

## Discussion

Based on our previous study of the $H_2O/\alpha$-$Fe_2O_3$(012)-(1×1) system[20], $H_2O$ exposure at 300 K should result in a (1×3) overlayer consisting of partially-dissociated water dimers ($H_2O$-$OH_t$). The experimental evidence for this was threefold: (1) temperature programmed desorption (TPD) data showed a desorption peak at 345 K containing 1.3 $D_2O$/unit cell, (2) XPS data showed approximately half the molecules were dissociated, (3) nc-AFM images showed 4 protrusions spread across 3 surface unit cells above the positions of the surface Fe cations.[20] The bimodal apparent height of the protrusions in nc-AFM was reproduced in simulations based on the DFT-determined partially-dissociated water dimer structure.[20] It is important to note, however, that the nc-AFM cannot be used to directly retrieve structural information about the species.

The primary result of this NIXSW study is that the oxygen atoms within the intact $H_2O$ and $OH_t$ species in fact have essentially identical adsorption heights above the $\alpha$-$Fe_2O_3$(012) surface, within experimental error ($H_{H2O} = 1.45 \pm 0.04$ Å and $H_{OHt} = 1.47 \pm 0.02$ Å). Similar adsorption heights imply similar Fe-O bond lengths, as all surface cation sites are equivalent within the bulk truncated $Fe_2O_3$(012) structure. Typically, NIXSW results cannot be trivially converted into a useful measure of bond length as the technique projects the positions onto the bulk crystal unit cell and is blind to any surface relaxations. Though here, our DFT calculations shows negligible vertical relaxation of the surface O and Fe for the surface after $H_2O$ exposure (Figure 1 and Table 1). Assuming bulk continuation lateral positions as was found in the nc-AFM images, the measured NIXSW heights give Fe-O bond lengths for both the $H_2O$ and $OH_t$ of approximately 2.0 Å.

This result is in contrast to previous quantitative studies of water adsorption on other metal oxides. Normally, metal-O bond lengths are found to be shorter ($\leq 1.9$ Å) for OH and longer ($\geq 2.1$ Å) for $H_2O$.[50,51,54,55] Reference calculations (HSE 12%) for isolated $H_2O$ and $OH_t$ on $\alpha$-$Fe_2O_3$(012) indeed yield bond lengths of 2.12 Å and 1.88 Å, respectively, in line with these expectations (Table 2). However, after the formation of the partially dissociated dimer, the Fe-$H_2O$ bond length shortens to 2.08 Å, while the Fe-$OH_t$ bond length extends to 1.94 Å (Table 2). This change can be understood in the context of cooperative binding interactions.[32,33] In the following, we emphasize the cooperative binding interactions observed in our study following the terminology and work of Schiros et al..[34] We refer the interested reader to their work for a more complete and detailed explanation.

The central concept in cooperative binding interactions for adsorbed molecules is the balance between surface bonding (S-bonding) and hydrogen bonding (H-bonding). The S-bonds and H-bonds can be either acceptor and/or donor bonds with regards to H-bond formation (note: this is not with regards to donating/ accepting charge). Exhibiting similar strengths, their equilibrium is crucial for the stability of molecules such as the $H_2O$-OH dimer observed here. In short, the more or the stronger the donor bonds received by a species, the stronger the acceptor bonds for that species.

For example, an adsorbed $H_2O$ molecule will form an S-bond with its adjacent surface Fe atom. From the $H_2O$ molecule's perspective this bond acts as an acceptor. In its isolated state, a $H_2O$ molecule does not form H-bonds via its H atom leading to the absence of donor bonds. If the molecule is instead binding to an adjacent OH molecule it will also form a donor H-bond via its H atom. To maintain balance, the strength of both bonds at the $H_2O$ molecule will scale with each other i.e. the existence or an increased strength of the donor H-bond leads to the strengthening of the acceptor S-bond. This phenomenon is cooperative binding, where energy is gained not only by the existence of the H-bond itself but also by its influence on the S-bond.

In the case of the $H_2O$ in the $H_2O$-OH dimer on the $\alpha$-$Fe_2O_3$(012) surface, the presence of a neighboring OH molecule and the resulting H-bond therefore leads to a strengthening of the Fe-$H_2O$ surface bond, resulting in a shorter bond length compared to the isolated $H_2O$ case. This is exactly what we observe experimentally. This enhancement is also confirmed by a charge accumulation in the Fe-$H_2O$ binding area in Fig. 5, indicating improved and stronger hybridization with the Fe atom, again compared to the isolated case.

As with an isolated $H_2O$, an isolated OH also forms an acceptor S-bond with its adjacent Fe surface atom and lacks any donor bonds. However, if it forms a hydrogen bond with a neighboring $H_2O$, from its perspective it acquires a second acceptor bond. In contrast to the $H_2O$ case, the two acceptor bonds end up competing with each other to maintain balance. Thus, in the case of the OH in the $H_2O$-OH dimer the Fe-OH surface bond is weakened from this competition. This is evident as a longer bond as we observe and a charge depletion in the Fe-OH area (see Fig. 5).

As well as the α-$Fe_2O_3$(012) surface, similar cooperative binding arguments have been invoked to explain the formation of partially dissociated dimers on other surfaces such as $Fe_3O_4$(001)[22], $Fe_3O_4$(111)[23,24], $RuO_2$(110)[25], Mg(100)[26], PdO(101)[27] and ZnO(1010)[28]. Scanning probe microscopy, temperature programmed desorption (TPD) and DFT were used to investigate the preferential formation of $H_2O$-OH dimers over isolated species, with these prior studies all concluding that such a phenomenon could be explained by the formation of favorable hydrogen bonds between the adsorbed species. For example, Haywood et. al. undertook a TPD/DFT study of $H_2O$ on PdO(101)[27] and came to the same conclusions; an isolated OH species has restricted H-bonding on PdO(101) and so $H_2O$-OH dimers with a favorable balance of H-bonding and S-bonding are required for dissociation. In the end, our results provide the first quantitative and direct structural evidence of this hydrogen bonding effect in these observed dimers.

## Conclusion

We have shown that the $H_2O$ and $OH_t$ of the (1×3) overlayer on α-$Fe_2O_3$(012)-(1×1) sit close to bulk continuation adsorption heights ($H_{H2O} = 1.45 \pm 0.04$ Å and $H_{OHt} = 1.47 \pm 0.02$ Å) corroborating our prior nc-AFM/DFT study.[20] We have also discerned the adsorption height of the $OH_s$ species located in the surface ($H_{OHs} = -0.01 \pm 0.02$ Å), which is essentially in plane with surface oxygen.

The $H_2O$ and $OH_t$ both sit essentially coplanar with similar Fe-O bond lengths (≈ 2.0 Å). This stands in contrast to prior studies of isolated $H_2O$ and OH on other metal oxides.[50,51,54,55] Typically, on other surfaces, the $H_2O$ bond length is found to be longer and the OH shorter than what was found in this study. We explain these unexpected Fe-O bond lengths by the formation of a hydrogen bond between the $H_2O$ and $OH_t$. In turn, this hydrogen bond effects the strengths of the Fe-O bonds via charge reorganization.[34] This is the first direct and quantitative measure of this cooperative binding effect, which was enabled by the formation of isolated $H_2O$-OH dimers on the α-$Fe_2O_3$(012)-(1×1) surface.

More broadly, these results emphasize the importance of considering $H_2O$-OH interactions on metal oxides surface. As seen, these interactions play a central role in defining the dissociative behavior of $H_2O$, which is an important phenomenon for applications of metal-oxides in catalysis. It is also central phenomenon for informing the acid/base behavior of metal-oxide surfaces, which is important in general mineralogy.

## Acknowledgements
GSP, PR, ARA, PS, ARA, CW and MM acknowledge funding from the European Research Council (ERC) under the European Union's Horizon 2020 research and innovation program (grant agreement No. [864628], Consolidator Research Grant 'E-SAC'). This work was also supported by the Austrian Science Fund (FWF) under project number F81, Taming Complexity in Materials Modeling (TACO) (MM, CW, GSP, UD, MS, MM, CF). The computational results presented have been achieved [in part] using the Vienna Scientific Cluster (VSC). We also thank Diamond Light Source for the award of beam time (SI31726-1)

# Supplementary Information
## Quantitative measurement of cooperative binding in partially-dissociated water dimers at the hematite "R-cut" surface

Paul T. P. Ryan[1], Panukorn Sombut[1], Ali Rafsanjani Abbasi[1], Chunlei Wang[1], Fulden Eratam[2], Francesco Goto[2,3], Ulrike Diebold[1], Matthias Meier[1,4], David A. Duncan[2], Gareth S. Parkinson[1]

[1]Institute of Applied Physics, Technische Universität Wien, Vienna, Austria
[2]Diamond Light Source, Harwell Science and Innovation Campus, Didcot, UK
[3]Politecnico di Milano, Piazza Leonardo da Vinci, Milano MI, Italy
[4]Faculty of Physics and Center for Computational Materials Science, University of Vienna, Vienna, Austria

# Contents

## 1. DFT further details

All the calculations were performed using the Vienna *ab initio* Simulation Package (VASP).[1,2] The Projector Augmented Wave (PAW) approach[3,4] was used for handling the near-core regions, with a basis set cut-off energy of 550 eV. Calculations were initially performed using the Perdew-Burke-Ernzerhof (PBE)[5] and revised PBE (revPBE)[6] exchange-correlation functional with an effective on-site Coulomb repulsion term $U_{eff}$ = 5.0 eV[7] for Fe atoms to model the oxide. Calculations are spin-polarized and performed at the Γ-point only for the (2 × 3) supercells and a Γ-centered k-mesh of 6×1×3 for the bulk optimization. Convergence is achieved when an electronic energy step of $10^{-6}$ eV is obtained, and forces acting on ions smaller than 0.02 eV/Å. Different van der Waals implementations were tested such as vdW corrections according to the method of Grimme et al. (DFT-D2)[8] with zero-damping (DFT-D3)[9] and the correction proposed by Dion *et al*. and Klimes *et al*. (vdW-DF, vdW-DF2, optPBE-DF, optB88-DF and optB86b-DF)[10–12]. We also utilized metaGGA functional (SCAN and R2SCAN) with the inclusion of vdW (rVV10)[13–16] and an on-site Coulomb repulsion term $U_{eff}$ = 3.10 eV.[17] We also tested a variant optimized specifically for solids, PBEsol,[18] which accounts for the well-known issue of disfavored density overlapping present in PBE, using the same $U_{eff}$. The hybrid functional (HSE06)[19] was investigated with the standard mixing factor 12% and 25% and screening length (0.2$^{-1}$ Å$^{-1}$). The supercell was reduced to (1 × 3) supercell with k-meshes 3×1×1. Symmetric slabs were built, consisting of four $Fe_4O_6$ layers in thickness where only the two inner central O layers are kept fixed. The bottom surface is saturated with a full monolayer of water molecules and left untouched throughout the study. A vacuum region of 15 Å between consecutive slabs normal to the surface is added to avoid interactions.

## 2. NIXSW OH position calculation

**Defining the models**

To calculate the individual $OH_t$ (OH species bound to a surface cation) and $OH_s$ (OH species formed from a lattice oxygen atom on the surface) coherent positions from the single NIXSW measurement of the $O_{OH}$ photoemission peak, a model of the OH distribution must be defined based on a set of assumptions about the possible distributions the OH species can take. Here we will outline all the models tested in this study including the model that is used in the main article.

All models are a two-atom model with each atom in the model representing a coherent location of one of the OH species ($OH_t$ and $OH_s$). In such a case, individual coherent positions, $p$, and coherent fractions, $f$, for each atom can be defined ($p_{OHt}$, $p_{OHs}$ and $f_{OHt}$, $f_{OHs}$). Note that for clarity the $hkl$ subscript denoting the reflection has been omitted and replaced with the species type. The coherent positions ($p_{OHt}$ and $p_{OHs}$) are unknown and to be determined. The coherent fractions ($f_{OHt}$ and $f_{OHs}$) are defined by the specific model and are the relative populations of the two sites. These coherent fractions must sum to unity: $f_{OHt} + f_{OHs} = 1$.

The measured coherent fraction, $f_m$, may be less than unity due to a specific distribution of atoms (as to be determined) but also due to molecular and crystal vibrations and contributions from decoherent species. These considerations lead to the following definition of $f_m$:

$$f_m = f_{str} \cdot D_H \cdot C \qquad (1)$$

where $f_{str}$ is the true structural coherent fraction due to the specific distribution of the two atoms alone.[20] $D_H$ is the Debye-Waller factor which defines how much the coherent fraction is reduced due to molecular and crystal vibrations. This kind of vibrational data is rare for specific species on specific surfaces. So for this study, $D_H$ of the OH species will take only two values; $D_H = 1.0$ i.e. no vibration or $D_H = 0.91$ which is the Debye-Waller factor measured in this study for H$_2$O. $C$ is an order parameter which defines the fraction of OH species that coherently contributes to the signal. For example, for the model used in the main article $C = 0.79$ from considering that all extra 21% OH contributes incoherently.

Six models with differing values of $D_H$, $C$ and $f_{OHt}$:$f_{OHs}$ ratio have been investigated. $p_{OHt}$ and $p_{OHs}$ for each of the models were calculated along with heights, $H_{OHt}$ and $H_{OHs}$, above a bulk truncated (012) oxygen surface. $H_{OHt}$ and $H_{OHs}$ were calculated according to equation 1 in the main article with values of $n = 1$ and $n = 0$ respectively. These results are presented in Table S1 along with the parameters defined. The next section outlines the details of the calculations while the final section discusses these results.

Table S1 – Results of all models tested in this study with model 2 being the model used in the main article. $H_{OHt}$ and $H_{OHs}$ have been calculated using equation 1 from the main article with $n = 1$ and $n = 0$ respectively. The errors in the $H_{OHt}$ and $H_{OHt}$ values are all $\pm 0.02$ Å.

| Model # | $f_{OHt}:f_{OHs}$ | $D_H$ | $C$ | $H_{OHt}$ (Å) | $H_{OHs}$ (Å) |
|---|---|---|---|---|---|
| 1 | 0.5:0.5 | 1 | 1 | 1.38 | 0.09 |
| 2 (main article) | 0.5:0.5 | 0.91 | 0.79 | 1.47 | -0.01 |
| 3 | 0.55:0.45 | 0.91 | 0.90 | 1.46 | 0.07 |
| 4 | 0.45:0.55 | 0.91 | 0.90 | 1.40 | 0.01 |
| 5 | 0.8:0.2 | 0.91 | 1 | 1.60 | 0.50 |
| 6 | 0.2:0.8 | 0.91 | 1 | 0.97 | -0.13 |

## Calculating $p_{OHt}$ and $p_{OHs}$

The structural information of each atom and the measurement results can each be represented as a vector in the complex plane (structure factors for the given reflection[20,21]) with values of $p$ and $f$ being the phase and magnitude, respectively, of each vector. The vectors of the two individual atom sites are $Z_{OHt} = f_{OHt}e^{2\pi i p_{OHt}}$ and $Z_{OHs} = f_{OHs}e^{2\pi i p_{OHs}}$ and their sum:

$$Z_m = Z_{OHt} + Z_{OHs} = f_{str}e^{2\pi i p_m} \tag{2}$$

produces a third vector, $Z_m$, who's $p_m$ and $f_{str}$ values are retrieved through the NIXSW measurement. The magnitudes of these vectors can be related to the difference in the phase, $\Delta p$, between $Z_{OHt}$ and $Z_{OHs}$ as such:

$$|Z_m|^2 = |Z_{OHt}|^2 + |Z_{OHs}|^2 + 2|Z_{OHt}||Z_{OHs}|\cos(2\pi\Delta p) \tag{3}$$

which can be rearranged and recast with respect to the vector magnitudes, $f_{str}$, $f_{OHt}$ and $f_{OHs}$:

$$\cos(2\pi\Delta p) = \frac{f_{str}^2 - f_{OHt}^2 - f_{OHs}^2}{2 f_{OHt} f_{OHs}} \tag{4}$$

The unit vectors $\hat{Z}_{OHt}$ and $\hat{Z}_{OHs}$ can then be defined dependent on $\Delta p$ and the measurement unit vector $\hat{Z}_m = e^{2\pi i p_m}$:

$$\hat{Z}_{OHt} = \frac{f_{OHt}}{f_{str}} e^{2\pi i p_m} + \frac{f_{OHs}}{f_{str}} e^{2\pi i (p_m - \Delta p)} \tag{5}$$

$$\hat{Z}_{OHs} = \frac{f_{OHs}}{f_{str}} e^{2\pi i p_m} + \frac{f_{OHt}}{f_{str}} e^{2\pi i (p_m + \Delta p)} \tag{6}$$

where the choice of plus or minus sign in $e^{2\pi i (p_m \pm \Delta p)}$ is such that the OH$_s$ has the larger $p_{OHs}$. Finally, the individual phases of the vectors, $p_{OHt}$ and $p_{OHs}$, are retrieved with $p = \frac{1}{2\pi} \tan^{-1}\left(\frac{Im(\hat{Z})}{Re(\hat{Z})}\right)$.

## Discussion of models

### Model 1

Model 1 (Table S1) calculates the heights not taking into account any molecular and crystal vibrations ($D_H = 1$) and assuming that all the OH species contribute coherently via the two atom sites ($C = 1$) by ignoring the excess 21% OH. This model places the OH$_t$ lower than the H$_2$O ($H_{OHt} = 1.38 \pm 0.02$ Å vs $H_{H2O} = 1.45 \pm 0.04$ Å), which is counter to the trend seen in all of the DFT results, though the significance of the difference is minor and typically a 0.1 Å difference between theory and experiment would be considered negligible.

**Model 2**

Model 2 includes the effect of both molecular and crystal vibrations and the possible decoherent effect of the excess OH. $D_H$ was set $D_H = 0.91$ which is the $D_H$ of H$_2$O on this surface and $C$ was set $C = 0.79$ from considering that all of the extra 21% OH species incoherently contribute. Including these affects in model 2 increases the separation between $p_{OHt}$ and $p_{OHs}$ and moves the OH$_t$ slightly higher than the H$_2$O as is observed in the DFT calculations. The height of the OH$_s$ is also still reasonable placing it in plane with the surface oxygens. This is the model that has been used in the main article.

**Model 3 and 4**

Models 3 and 4 place half of the excess OH coherently in either the OH$_t$ or the OH$_s$ sites respectively. This simulates a partial coherent contribution from the defect sites. The other half of the excess OH is considered incoherently contributing via the order parameter $C = 0.90$. In both models, $D_H = 0.91$ as for model 2 and H$_2$O. The $H_{OHt}$ in models 3 and 2 are the same within the experimental error. However, model 3 places the OH$_s$ more above the O surface plane (in contrast to model 2) which is consistent with the DFT. It is not unreasonable to think that any excess OH may be due to water adsorption at surface defects which contribute somewhat coherently to the measurement and model 4 could be reflecting this, similar to what is observed for water on TiO$_2$(110).[22]

**Model 5 and 6**

Finally, models 5 and 6 place all of the excess OH coherently in the OH$_t$ or the OH$_s$ sites respectively ($C = 1$ and $D_H = 1$ in both cases). These models are unlikely, given the prior AFM/ XPS/ DFT study and are provided here as extreme cases. These models either place the OH$_s$ physically too high ($H_{OHs} = 0.50 \pm 0.02$ Å for model 5) or the OH$_t$ too low ($H_{OHt} = 0.97 \pm 0.02$ Å for model 6) and are thus unlikely. Models with higher relative populations at either site (> 0.8) are incalculable if $D_H$ and $C$ are to remain fairly high (i.e. $D_H \geq 0.9$ and $C \geq 0.79$). Though $D_H < 0.9$ would require unreasonably large molecular and crystal vibrational amplitudes and $C < 0.79$ would require the observation of more defects.

## 3. Comparison of all DFT functionals

**Optimized vs experimental lattice parameters**

Table S2 shows the DFT results for all functionals tested using the experimental lattice parameters (a = 5.038 Å and c = 13.77 Å). However, bulk structure calculations can be carried out for each functional and so called optimized lattice parameters can be retrieved from the resulting relaxed bulk structure. Generally, these optimized parameters are different to the experimental lattice parameters. This is depicted in Figure S1 which shows for each functional the difference between the optimized and experimental $a$ (Figure S1 a) and $c$ (Figures S1 b) lattice parameters. Generally speaking, the hybrid calculations perform the best in reproducing the experimental lattice parameters, the strongest deviations from experiment are found for functionals with dispersion corrections and the SCAN functionals all underpredict the lattice parameters.

The question arises as to which set of lattice parameters to use; optimized vs experimental. As such, a second set of calculations of the surface were undertaken but instead using separate lattice parameters from optimized bulk structure calculations for each functional. These results are provided in Table S3. However, the direct comparison of the absolute heights between the optimized and experimental lattice parameters is not possible. This is because changing the lattice parameters changes the reflection layer spacing ($d_{hkl}$) rendering any direct comparison of the absolute DFT heights with the NIXSW heights impossible. Moreover, a deviation from the experimental lattice parameters represents a relaxation of the structure and this would work to change the projected height of adsorbed species while not necessarily changing the height with respect to the oxygen surface layer. As such, instead of the absolute heights, the coherent positions, $p_{024}$, must be compared between optimized and experimental lattice parameters. In this way, the coherent positions are normalized heights with respect to $d_{hkl}$ and this therefore removes the effects of any changes to $d_{hkl}$.

Figure S2 shows a comparison between the $p_{024}$ of the optimized and experimental lattice parameter for the H$_2$O molecule in the H$_2$O-OH$_t$ dimer. When using the experimental lattice parameters, a number of functionals give values close to the NIXSW $p_{024} = 0.79$ value (with respect to a projected bulk oxygen surface layer at $p_{024} = 0.61$). The best performing are the PBE, PBE-D3, optB88-DF and HSE 12% (Table S2). However, when using the optimized lattice parameters, all the functionals over bind the position of the H$_2$O with none being able to reproduce the NIXSW $p_{024} = 0.79$ value. The highest reached $p_{024}$ are for the revPBE and vdw-DF functionals ($p_{024} = 0.76$ and $0.77$) though this is likely due to their very large overprediction of the lattice parameters ($\Delta c = +0.18$ Å and $+0.23$ Å).

The Fe-O bond lengths for the H$_2$O and OH$_t$ species will also affect the calculated heights. Figure S3 shows the bond lengths for the H$_2$O ($d_{Fe-H2O}$) and OH$_t$ ($d_{Fe-OHt}$) species when calculated using the optimized (Figure S3 a and b) and experimental (Figure S3 c and d) lattice parameters. The bond lengths are essentially the same in each case and both follow the same trend as in Figure S1 for the deviation from experiment. This makes clear that there are negligible differences in local effects between the optimized and experimental lattice parameters and any differences are almost entirely from bulk and/or surface relaxations.

To further quantify the correlation of the bulk and surface structural relaxations with the relevant structural values (e.g. heights and bond lengths), correlation coefficients were calculated for all parameters in Tables S1 and S2 with respect to $\Delta c$ as well as for the height difference, $\Delta H$, between the H$_2$O and OH$_t$ species. Table S3 provides these coefficients. In both the optimized and experimental lattice parameter cases, all heights and bond lengths show extremely strong correlation with $\Delta c$. Slightly weaker correlation is found for Fe-OH bond lengths (coefficient > 0.6) compared to Fe-H$_2$O bond lengths (coefficient > 0.9) and this is likely due to the more ionic character of the Fe-OH bond. For the optimized case, the OH$_s$ height shows no correlation with $\Delta c$ (coefficient = 0.04) but this can be attributed to the OH$_s$ sitting at a bulk position who's coherent position would be consistent. The only parameter showing little or no correlation with $\Delta c$ for both the optimized and experimental lattice parameter cases is $\Delta H$. $\Delta H$ can be thought of as a measure of the cooperativity effect. This lack of correlation is this likely due to differences in local effects defining the strength of the cooperativity between each functional.

In light of all of the above, the experimental lattice parameters are used for comparison of the DFT to the NIXSW results and will be used throughout.

Table S2 – **DFT calculations using experimental lattice parameters (a = 5.038 Å, c = 13.77 Å).** The DFT heights ($H_{H2O}$, $H_{OHt}$ and $H_{OHs}$), with respect to an oxygen bulk terminated (012) surface for each functional with comparison to the NIXSW results. Values in brackets for the NIXSW results are the error in the last significant figure. $d_{Fe-H2O}$ and $d_{Fe-OHt}$ are the relevant bonds lengths to surface Fe cations. $d_{Fe-H2O(iso)}$ and $d_{Fe-OHt(iso)}$ are the bond lengths for isolated species. $p_{024}^{H2O}$ is the coherent position of the H$_2$O. $\Delta c$ is the difference between the experimental and calculated c unit cell parameter for α-Fe$_2$O$_3$. The green highlighted rows show the results presented in the main article.

| Functional | $U_{eff}$ (eV) | $H_{H2O}$ (Å) | $H_{OHt}$ (Å) | $H_{OHs}$ (Å) | $d_{Fe-H2O}$ (Å) | $d_{Fe-OHt}$ (Å) | $d_{Fe-H2O(iso)}$ (Å) | $d_{Fe-OHt(iso)}$ (Å) | $p_{024}^{H2O}$ | $\Delta c$ (Å) |
|---|---|---|---|---|---|---|---|---|---|---|
| NIXSW | n/a | 1.45(4) | 1.47(2) | -0.01(2) | n/a | n/a | n/a | n/a | n/a | n/a |
| PBE | 5 | 1.47 | 1.52 | 0.16 | 2.07 | 1.96 | 2.11 | 1.90 | 0.80 | 0.08 |
| PBE-D2 | | 1.38 | 1.45 | 0.10 | 2.06 | 1.96 | 2.10 | 1.90 | 0.75 | -0.10 |
| PBE-D3 | | 1.43 | 1.49 | 0.13 | 2.07 | 1.95 | 2.11 | 1.90 | 0.78 | 0.03 |
| vdw-DF | | 1.69 | 1.75 | 0.32 | 2.13 | 1.96 | 2.16 | 1.91 | 0.92 | 0.23 |
| vdW-DF2 | | 1.77 | 1.84 | 0.40 | 2.12 | 1.96 | 2.15 | 1.91 | 0.96 | 0.31 |
| optPBE-DF | | 1.52 | 1.57 | 0.18 | 2.09 | 1.96 | 2.12 | 1.90 | 0.83 | 0.10 |
| optB88-DF | | 1.42 | 1.47 | 0.11 | 2.07 | 1.95 | 2.10 | 1.90 | 0.77 | 0.02 |
| optB86b-DF | | 1.38 | 1.42 | 0.07 | 2.06 | 1.95 | 2.09 | 1.90 | 0.75 | -0.02 |
| SCAN | | 1.30 | 1.35 | 0.00 | 2.04 | 1.94 | 2.07 | 1.88 | 0.70 | -0.10 |
| SCAN+rVV10 | | 1.28 | 1.34 | 0.01 | 2.03 | 1.93 | 2.07 | 1.89 | 0.69 | -0.12 |
| R2SCAN | | 1.32 | 1.37 | 0.03 | 2.04 | 1.94 | 2.09 | 1.89 | 0.72 | -0.07 |
| R2SCAN+rVV10 | | 1.29 | 1.34 | 0.00 | 2.04 | 1.94 | 2.08 | 1.89 | 0.70 | -0.11 |
| revPBE | | 1.63 | 1.68 | 0.27 | 2.11 | 1.96 | 2.15 | 1.91 | 0.88 | 0.18 |
| PBEsol | | 1.27 | 1.32 | 0.00 | 2.03 | 1.95 | 2.06 | 1.90 | 0.69 | -0.10 |
| HSE06(25%) | n/a | 1.36 | 1.44 | 0.06 | 2.07 | 1.93 | 2.11 | 1.88 | 0.74 | -0.01 |
| HSE06(12%) | | 1.42 | 1.50 | 0.11 | 2.08 | 1.94 | 2.12 | 1.88 | 0.77 | 0.06 |
| SCAN | 3.1 | 1.30 | 1.35 | 0.01 | 2.04 | 1.94 | 2.08 | 1.88 | 0.71 | -0.06 |
| SCAN+rVV10 | | 1.28 | 1.34 | 0.01 | 2.04 | 1.94 | 2.08 | 1.88 | 0.70 | -0.09 |
| R2SCAN | | 1.34 | 1.41 | 0.05 | 2.05 | 1.94 | 2.09 | 1.89 | 0.73 | -0.03 |
| R2SCAN+rVV10 | | 1.30 | 1.35 | 0.01 | 2.05 | 1.94 | 2.08 | 1.89 | 0.71 | -0.06 |

Table S3 - **DFT calculations using optimized lattice parameters**. The DFT heights ($H_{H2O}$, $H_{OHt}$ and $H_{OHs}$), with respect to an oxygen bulk terminated (012) surface for each functional with comparison to the NIXSW results. Values in brackets for the NIXSW results are the error in the last significant figure. $d_{Fe-H2O}$ and $d_{Fe-OHt}$ are the relevant bonds lengths to surface Fe cations. $d_{Fe-H2O(iso)}$ and $d_{Fe-OHt(iso)}$ are the bond lengths for isolated species. $p_{024}^{H2O}$ is the coherent position of the H$_2$O. $\Delta c$ is the difference between the experimental and calculated c unit cell parameter for α-Fe$_2$O$_3$.

| Functional | $U_{eff}$ (eV) | $H_{H2O}$ (Å) | $H_{OHt}$ (Å) | $H_{OHs}$ (Å) | $d_{Fe-H2O}$ (Å) | $d_{Fe-OHt}$ (Å) | $d_{Fe-H2O(iso)}$ (Å) | $d_{Fe-OHt(iso)}$ (Å) | $p_{024}^{H2O}$ | $\Delta c$ (Å) |
|---|---|---|---|---|---|---|---|---|---|---|
| NIXSW | n/a | 1.45(4) | 1.47(2) | -0.01(2) | n/a | n/a | n/a | n/a | 0.79 | n/a |
| PBE | 5 | 1.38 | 1.43 | 0.08 | 2.08 | 1.96 | 2.11 | 1.90 | 0.75 | 0.08 |
| PBE-D2 | 5 | 1.36 | 1.43 | 0.09 | 2.06 | 1.95 | 2.10 | 1.90 | 0.74 | -0.10 |
| PBE-D3 | 5 | 1.38 | 1.44 | 0.08 | 2.07 | 1.96 | 2.11 | 1.90 | 0.75 | 0.03 |
| vdw-DF | 5 | 1.44 | 1.48 | 0.07 | 2.13 | 1.96 | 2.17 | 1.91 | 0.77 | 0.23 |
| vdW-DF2 | 5 | 1.43 | 1.48 | 0.07 | 2.13 | 1.96 | 2.17 | 1.91 | 0.76 | 0.31 |
| optPBE-DF | 5 | 1.40 | 1.45 | 0.07 | 2.09 | 1.96 | 2.13 | 1.90 | 0.75 | 0.10 |
| optB88-DF | 5 | 1.37 | 1.42 | 0.06 | 2.07 | 1.95 | 2.10 | 1.90 | 0.74 | 0.02 |
| optB86b-DF | 5 | 1.36 | 1.41 | 0.06 | 2.06 | 1.95 | 2.09 | 1.90 | 0.74 | -0.02 |
| SCAN | 5 | 1.36 | 1.40 | 0.06 | 2.03 | 1.94 | 2.07 | 1.89 | 0.74 | -0.10 |
| SCAN+rVV10 | 5 | 1.34 | 1.39 | 0.06 | 2.03 | 1.94 | 2.06 | 1.89 | 0.73 | -0.12 |
| R2SCAN | 5 | 1.36 | 1.40 | 0.06 | 2.04 | 1.94 | 2.08 | 1.89 | 0.74 | -0.07 |
| R2SCAN+rVV10 | 5 | 1.35 | 1.40 | 0.06 | 2.04 | 1.94 | 2.08 | 1.89 | 0.74 | -0.11 |
| revPBE | 5 | 1.42 | 1.47 | 0.07 | 2.12 | 1.96 | 2.16 | 1.90 | 0.76 | 0.18 |
| PBEsol | 5 | 1.34 | 1.39 | 0.07 | 2.03 | 1.95 | 2.06 | 1.90 | 0.73 | -0.10 |
| HSE06(25%) | n/a | 1.36 | 1.44 | 0.06 | 2.07 | 1.93 | 2.11 | 1.88 | 0.74 | -0.01 |
| HSE06(12%) | n/a | 1.37 | 1.46 | 0.07 | 2.08 | 1.94 | 2.12 | 1.88 | 0.74 | 0.06 |
| SCAN | 3.1 | 1.36 | 1.42 | 0.07 | 2.04 | 1.94 | 2.07 | 1.89 | 0.74 | -0.06 |
| SCAN+rVV10 | 3.1 | 1.36 | 1.42 | 0.07 | 2.04 | 1.94 | 2.07 | 1.89 | 0.74 | -0.09 |
| R2SCAN | 3.1 | 1.36 | 1.42 | 0.07 | 2.05 | 1.94 | 2.09 | 1.89 | 0.74 | -0.03 |
| R2SCAN+rVV10 | 3.1 | 1.36 | 1.42 | 0.07 | 2.05 | 1.94 | 2.08 | 1.89 | 0.74 | -0.06 |

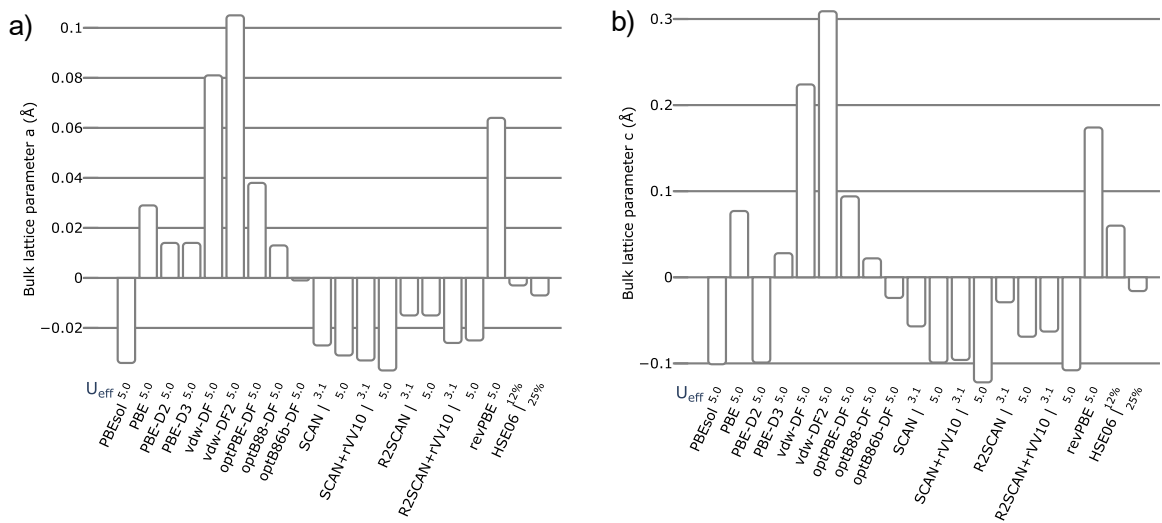

Figure S1 – The difference between the experimental a) bulk a lattice parameter and b) bulk c lattice parameter with that of an optimized bulk structure calculation for each functional.

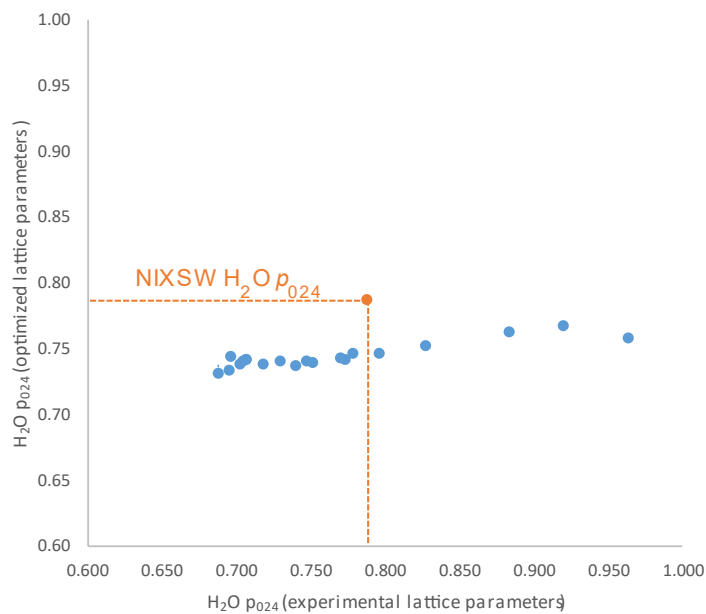

Figure S2 – The coherent positions, $p_{024}$, of the $H_2O$ for calculations using the optimized vs experimental lattice parameters. Only when using the experimental lattice parameters can the DFT reproduce the NIXSW structural results. And only those functionals that have small bulk or surface relaxations are suitable for the structure calculations.

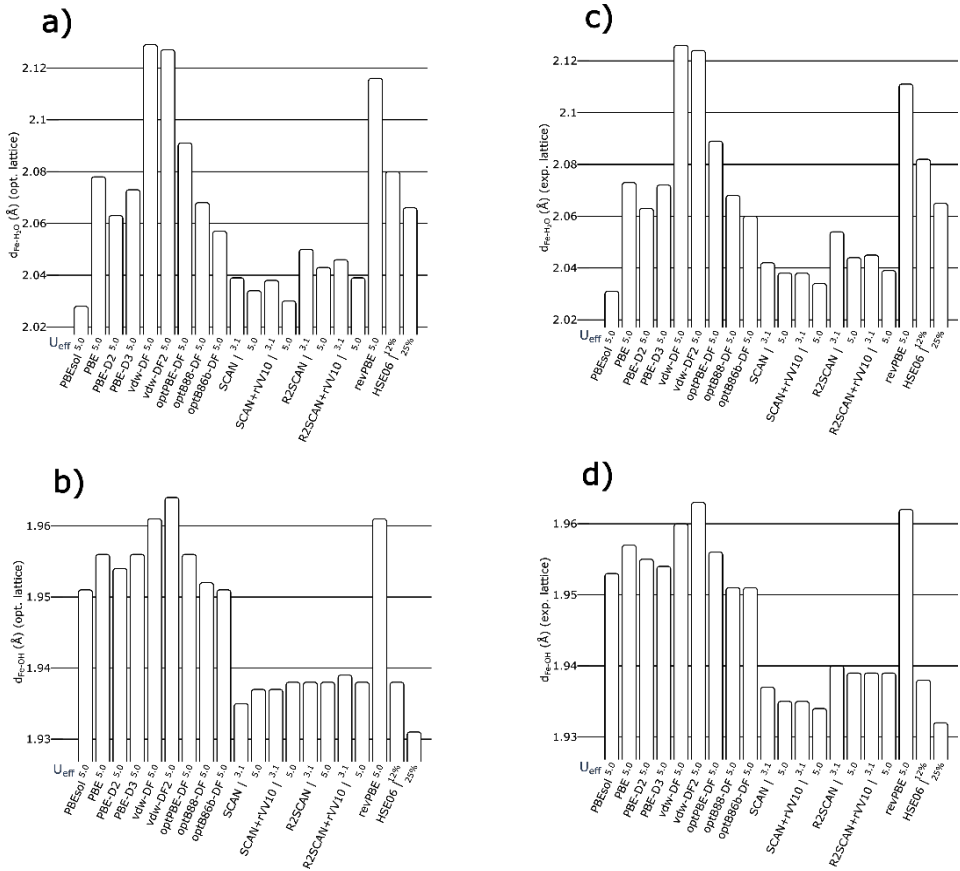

Figure S3 – The Fe-O bond lengths for the $H_2O$ and $OH_t$ species using the optimized (a and b) and experimental (c and d) lattice parameters. The trends across the functionals are identical between optimized vs experimental and correlate with Figure S1.

Table S3 – Calculated correlation coefficients for $\Delta c$ with respect to the heights ($H$), bond lengths ($d$), $H_2O$ coherent positions ($p_{024}^{H2O}$) and differences between the $H_2O$ and $OH_t$ heights ($\Delta H$). Both optimized and experimental lattice parameter cases are provided. For all heights and bond lengths there is extremely strong correlation in both cases. For $\Delta H$ there is very little correlation for either the optimised or experimental cases. This is likely due to differences in local effects defining the strength of the cooperativity between each functional.

|  | $H_{H2O}$ | $H_{OHt}$ | $H_{OHs}$ | $d_{Fe-H2O}$ | $d_{Fe-OHt}$ | $d_{Fe-H2O(iso)}$ | $d_{Fe-OHt(iso)}$ | $p_{024}^{H2O}$ | $\Delta H$ |
|---|---|---|---|---|---|---|---|---|---|
| optimised | 0.94 | 0.91 | 0.04 | 0.96 | 0.73 | 0.96 | 0.60 | 0.88 | -0.15 |
| experimental | 0.98 | 0.98 | 0.97 | 0.96 | 0.72 | 0.95 | 0.62 | 0.98 | 0.17 |

## Comparison of functionals

Figure S4 shows the difference between the DFT and NIXSW heights for each functional using the experimental lattice parameters for the $H_2O$, $OH_t$ and $OH_s$ species. Unsurprisingly, across all functionals there is a strong correlation with Figure S1, as the bulk and/or surface relaxations will directly affect the calculated heights and there is a direct correlation of this with the Fe-O bond lengths (Figures S3 and Table S3). Taking into account all three species, the best performing functionals are: PBE, PBE-D3 optB88-DF and HSE12%.

While the PBE functional closely reproduces the NIXSW heights, this is likely a coincidence due to a strong relaxation of the unit cell outward, normal to the surface. This is evident from the large deviation of its $c$ lattice parameter ($\Delta c = +0.08$ Å). As such, the best performing functionals which take into account bulk and surface relaxations are the PBE-D3, optB88-DF and HSE 12% calculations as these all have very low deviations from the experimental lattice parameters.

While reproducing the absolute values of the NIXSW results, the best performing functional should also reproduce observed experimental trends. It is clear that the NIXSW results demonstrate that the $H_2O$ and $OH_t$ sit very close in height to one another and this can generally be thought of as a direct consequence of the cooperativity effect. More specifically, all reasonable distribution Models for the NIXSW results place the $OH_t$ at essentially the same height as the $H_2O$ in the range -0.07 – +0.02 Å.

Figure S5 shows the combined relative differences between the $H_2O$ and $OH_t$ heights for each functional using the experimental lattice parameters. It is clear from Figure S4 that the PBE-D3 and hybrid calculations perform poorly, consistently placing the $OH_t$ much too high compared to the $H_2O$ (+ 0.08 Å). The best functional in this regard is optB88-DF which places the $H_2O$ and $OH_t$ close in height (+ 0.04 Å).

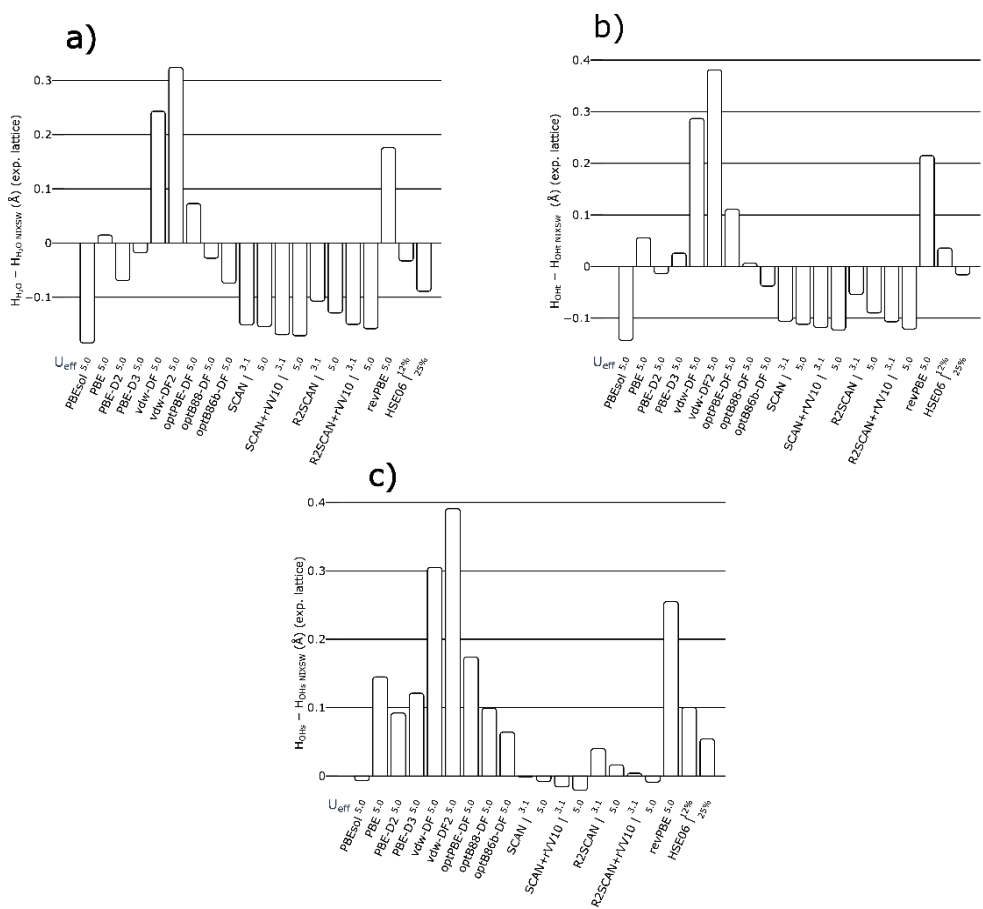

Figure S4 – The difference between the DFT and NIXSW heights for the a) $H_2O$, b) $OH_t$ and c) $OH_s$ species.

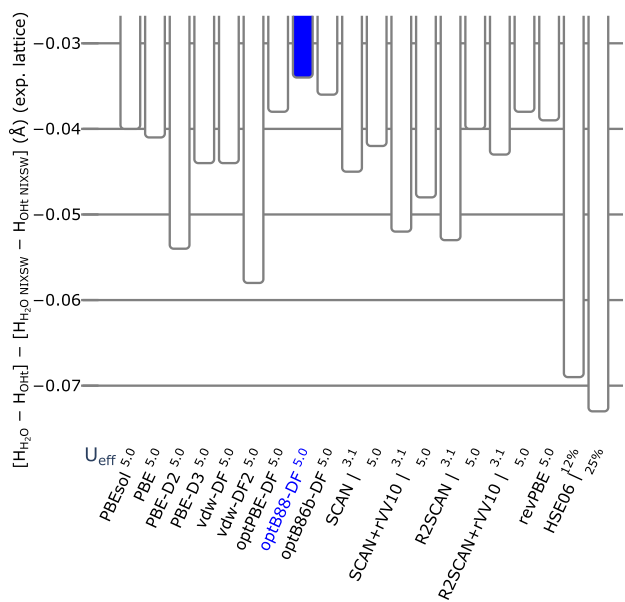

Figure S5 – The combined relative height differences between the H$_2$O and OH$_t$ species. Marked in blue is the optB88-DF functional, which in this regards performs the best, placing the H$_2$O and OH$_t$ species very close in height.

## Observation of cooperativity in DFT

Figure S6 shows the difference between the bond lengths of H$_2$O and OH$_t$ in the H$_2$O-OH$_t$ dimer vs for the isolated species (using experimental lattice parameters). This demonstrates directly the strength of the cooperativity effect. All functionals show some amount of cooperativity, with the H$_2$O bond being shorter and the OH$_t$ bond being longer in the dimer in all cases. However, there are large differences in the strength of the cooperativity between the functionals. While the optB88-DF performs well in reproducing the NIXSW results, it gives a fairly mediocre cooperativity effect. Stronger cooperativity is seen in the hydrid and revPBE calculations. This shows that, while the cooperativity effect plays a role in the similar measured adsorption heights of the H$_2$O and the OH$_t$, other effects, such as specific surface atom relaxation, would also play a role. Though such effects become harder to experimentally pinpoint.

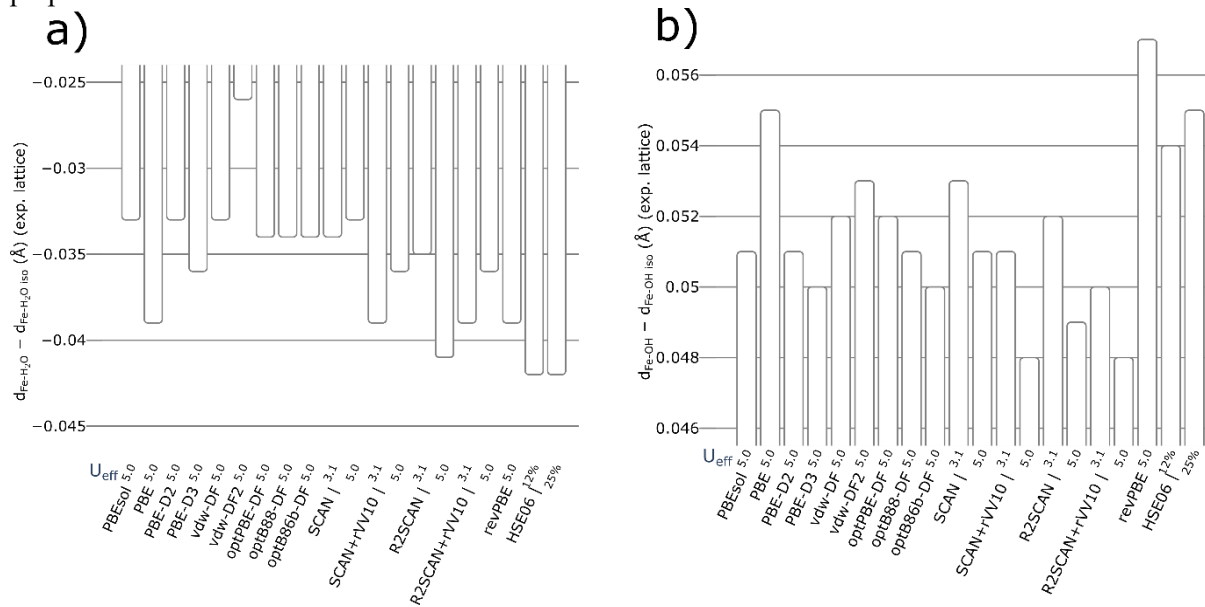

Figure S6 – The difference between the Fe-O bond lengths for isolated molecules vs molecules found in the H$_2$O-OH$_t$ dimer for the a) H$_2$O and b) OH$_t$ species. These DFT calculations utilized the experimental lattice parameters. This figure directly depicts the cooperativity effect. All functionals show a cooperativity effect.